\documentstyle[12pt,epsf,epsfig]{article}

\newcommand{\beq}{ \begin{eqnarray} }
\newcommand{\eeq}{ \end{eqnarray} }
\newcommand{\beqstar}{ \begin{eqnarray*} }
\newcommand{\eeqstar}{ \end{eqnarray*} }
\newcommand{\gsim}{ \mathop{}_{\textstyle \sim}^{\textstyle >} }
\newcommand{\lsim}{ \mathop{}_{\textstyle \sim}^{\textstyle <} }

\begin{document}
\baselineskip 0.7cm

\begin{titlepage}

\begin{center}

\hfill \today

{\large 
Revisiting Signature of Minimal Gauge Mediation
}
\vspace{1cm}

{\bf Junji Hisano}$^{1}$ and 
{\bf Yasuhiro Shimizu}$^{2}$
\vskip 0.15in
{\it
$^1${ICRR, University of Tokyo, Kashiwa 277-8582, Japan }\\
$^2${ Department of Physics, Korea Advanced Institute of Science and
Technology, Daejeon 305-701, Korea 
}\\
}
\vskip 0.5in

\abstract{ 

  We revisit phenomenology of the minimal gauge-mediated model. This
  model is motivated from the SUSY CP and flavor problems. A specific
  feature of this model is that $\tan\beta$ is naturally large, since
  the $B$ term in the Higgs potential is zero at the messenger
  scale. This leads to significant SUSY contributions to various
  low-energy observables.  We evaluate the anomalous magnetic moment
  of the muon and the branching ratio of $\overline{B}\to X_s\gamma$
  taking account of recent theoretical and experimental developments.
  We find that the current experimental data prefer a low messenger
  scale ($\sim 100$~TeV) and gluino mass around 1~TeV.  We also
  calculate the branching ratios of $\overline{B}\to X_s l^+l^-$,
  $B_s\to \mu^+\mu^-$, and $B^-\to \tau^-\,\overline{\nu}$, and show
  that these observables are strongly correlated with each other in
  this model.  }
\end{center}
\end{titlepage}
\section{Introduction}

Low-energy supersymmetry (SUSY) is a very attractive model of physics
beyond the standard model (SM). In the minimal supersymmetric standard
model (MSSM), however, general SUSY breaking masses of squarks and
sleptons induce too large FCNC and/or CP violation effects in
low-energy observables. These SUSY FCNC and CP problems should be
solved in realistic SUSY breaking models.
 
Gauge-mediated SUSY breaking
\cite{Dine:1981za,Dine:1993yw,Dine:1994vc,Dine:1995ag} is one of the promising
mechanisms to describe the SUSY breaking sector in the MSSM.  The SUSY
breaking is transmitted to the MSSM sector through the gauge
interaction, which induces the flavor-blind soft SUSY breaking masses
of squarks and sleptons. In the minimal gauge-mediated model (MGM)
\cite{Dine:1996xk,Rattazzi:1996fb}, the trilinear scalar couplings ($A$
terms) and Higgs bilinear coupling ($B$ term) are zero at tree level,
and they are induced from radiative corrections of the gaugino masses.
In this case, dangerous SUSY CP phases are absent, and the SUSY CP
problem is solved.

In Refs.~\cite{Gabrielli:1997jp,Gabrielli:1998sw}, phenomenology of
the MGM was studied.  One of the specific features in the MGM is that
$\tan\beta$ is naturally large, and significant SUSY contributions are
expected in various low-energy observables. They considered anomalous
magnetic moment of the muon, the $\overline{B}\to X_s\gamma$ decay, and
the $\overline{B}\to X_s l^+l^-$ decay. It was shown that the
deviations of ${\cal B}(\overline{B}\to X_s\gamma)$ and anomalous
magnetic moment of the muon from the SM predictions are strongly
correlated in the MGM.

Recently, theoretical calculations in both ${\cal B}(\overline{B}\to
X_s\gamma)$ and anomalous magnetic moment of the muon have been
improved. Also the experimental data was updated.  According to
Refs.~\cite{Misiak:2006zs,Becher:2006pu}, a theoretical estimation of
${\cal B}(\overline{B}\to X_s\gamma)$ at next-to-next-to-leading order
(NNLO) is lower than the experimental world average at a 1.4 $\sigma$
level.  In Refs.~\cite{Hagiwara:2006jt,Miller:2007kk} the SM
prediction of anomalous magnetic moment of the muon has been updated
by using the recent $e^+e^-\to\pi^+\pi^-$ data.  The new SM prediction
is larger than the experimental data at a 3.4 $\sigma$ level.  While
we can not still conclude that the deviations come from the new
physics, there is a room for additional new physics contributions.

In this paper, we revisit phenomenology of the MGM taking account of
the recent developments of ${\cal B}(\overline{B}\to X_s\gamma)$ and
anomalous magnetic moment of the muon. Also, we evaluate ${\cal
  B}(\overline{B}\to X_s l^+l^-)$, ${\cal B}(B_s\to \mu^+\mu^-)$, and
${\cal B}(B^-\to \tau^-\,\overline{\nu})$. The SUSY contributions to all the
observables are enhanced when $\tan\beta$ is large. We show that
the deviations from the SM predictions are strongly correlated in the
MGM and the recent result of ${\cal B}(\overline{B}\to X_s\gamma)$
prefers a low messenger scale and gluino mass around 1~TeV. 
This model will be tested by various low-energy experiments in
addition to the SUSY direct search at LHC.

\section{Minimal Gauge-Mediated Model}
\label{MGM}
First, let us briefly review the MGM. In the gauge-mediated SUSY
breaking, the SUSY breaking is mediated to the MSSM sector through the
SM gauge interactions. In the MGM, $N$ pairs of SU(2)$_L$ doublet and
of SU(3)$_C$ triplet chiral superfields are introduced as messenger
fields.  The SU(2)$_L$ doublets and of SU(3)$_C$ triplets are assumed
to be in SU(5) $\bf 5$ and $\bf 5^\star$-dimensional multiplets.
The messenger fields couple to a singlet chiral field by the
following superpotential,
\begin{equation}
 W = \lambda_i S \Phi_i \overline{\Phi}_i,
\end{equation}
where $S$ is a singlet, $\Phi_{2(3)}$ and $\overline{\Phi}_{2(3)}$ are
SU(2)$_L$-doublet (SU(3)$_C$-triplet) superfields. The scalar
component of $S$ develops a vacuum expectation value ($\langle
S\rangle$), which gives supersymmetric mass terms to the messenger
fields, $M_{M_i}=\lambda_i \langle S\rangle$.  Also the non-vanishing
$F$-component of $S$ $(\langle F_S \rangle)$ induces the SUSY breaking
in the messenger fields.

The gaugino masses in the MSSM are generated by one-loop diagrams of
the messenger fields and given by
\begin{eqnarray}
 \overline{M}_i =N \frac{\alpha_i}{4\pi} \Lambda g(x_i)\equiv \hat{M}_i g(x_i),
\end{eqnarray}
where $\Lambda=\langle F_S \rangle/\langle S \rangle$ and
$x_i=\Lambda/M_{M_i}$. The loop function $g(x_i)$ is given in Ref
\cite{Martin:1996zb}, but we take $x_i\lsim 1$ in this paper so that
$g(x_i) \simeq 1$. Hereafter, the parameters at the messenger scale
$M_M$ are denoted by bar.  The sfermion mass terms are generated by
two-loop diagrams and given by
\begin{eqnarray}
 \overline{m}_\alpha^2 
  \simeq \frac{1}{N}\left(2 C_3 \hat{M}^2_3 + 2 C_2 \hat{M}^2_2
+ \frac{6}{5}Y^2 \hat{M}_1^2 \right),
\end{eqnarray}
when $x_i\lsim 1$. Here, $C_i$ and $Y$ are the quadratic Casimir and
the hypercharge of sfermions, respectively. In the gauge-mediated
SUSY breaking, the squarks become rather heavy compared with the
sleptons and Higgs bosons since the soft SUSY breaking terms are
proportional to the gauge coupling constants.  In the MGM the
trilinear scalar couplings ($A$) and the bilinear Higgs coupling ($B$)
are assumed to vanish at the messenger scale,
\begin{equation}
 \overline{A}=\overline{B}=0.
\end{equation} 
Those $A$ and $B$ terms are induced by the gaugino-loop diagrams at the
weak scale, and the relative phases between the gaugino masses and the
$A$ and $B$ terms vanish, arg($M_i A^*$)=arg($M_i B^*$)=0.  Therefore, the
dangerous SUSY CP phases are absent, and the SUSY CP problem is solved
naturally in the MGM.

With the boundary conditions at the messenger scale, we can evaluate
the SUSY breaking parameters at the weak scale by solving the
renormalization group equations (RGEs).  For the $\mu$ parameter we do
not specify its origin, and it is determined by imposing the correct
electroweak symmetry breaking at the weak scale, as 
\begin{equation}
 \mu^2 = \lambda_t^2 \Delta_t^2 -m_H^2-\frac{1}{2}m_Z^2 (1+\delta_H),
\end{equation}
where $\lambda_t$ is the top-quark Yukawa coupling,  $m_H$ is the
common Higgs boson mass, and $\Delta_t^2$ and  $\delta_H$ represent
radiative corrections defined in Ref.~\cite{Rattazzi:1996fb}.

The $B$ term is generated radiatively at the weak scale.  It has two
types of contributions: one is a Higgsino-gaugino contribution $(B_G)$
and the other is an effective $A$-term contribution $(B_A)$.  They are
destructive to each other, and the magnitude of the $B$ parameter is
even smaller than one naively expected.  We follow the result in
Ref.~\cite{Rattazzi:1996fb} for the numerical calculation of the $B$
parameter.  Once the $B$ parameter is fixed, $\tan\beta$ is determined
by the minimization condition of the Higgs potential,
\begin{eqnarray}
\label{tanb}
 \tan\beta=-\frac{m_{A}^2}{B\mu},
\end{eqnarray}
where $m_{A}$ is the CP-odd Higgs boson mass.  

Notice that $\tan\beta$ is not a free parameter in the MGM, but, is
predicted from other parameters. Since the $B$ parameter is generated
radiatively, $\tan\beta$ becomes naturally large.  Also, sign$(B)$ is
opposite to sign$(M_2)$ in most parameter regions.  Then
sign$(M_2\mu\tan\beta)$ is determined to be positive from
Eq.~(\ref{tanb}). Moreover, in the MGM, all the soft SUSY breaking
parameters are determined by only three input parameters, $\Lambda$,
$M_M$, and $N$.  These are specific features of the MGM and very
important when we consider phenomenology of the MGM.

\section{Low-Energy Observables}

\subsection{Anomalous Magnetic Moment of the Muon}
\label{amu}
The current experimental measurement of the anomalous magnetic moment 
of the muon, $a_\mu=(g-2)_\mu/2$, was reported by Muon (g-2)
Collaboration \cite{Bennett:2006fi},
\begin{equation}
 a_\mu^{\rm exp}=11659208.0(6.3) \times 10^{-10}.
\end{equation}
Recently, the SM prediction of $a_\mu$ has been updated based on the
new data of $e^+e^-\to\pi^+\pi^-$. The most recent calculation
including new evaluation of the hadronic light-by-light scattering
contribution is \cite{Hagiwara:2006jt,Miller:2007kk}
\begin{equation}
  a_\mu^{\rm SM}=11659178.5(6.1) \times 10^{-10}.
\end{equation}
The discrepancy between the experimental data and the SM prediction is
\begin{equation}
\label{deltamu}
 \delta a_\mu \equiv a_\mu^{\rm exp}-a_\mu^{\rm SM}
= (29.5\pm 8.8)\times 10^{-10}.
\end{equation}
The most recent result indicates a 3.4$\sigma$ deviation, and it suggests 
new physics contributions.

In the MGM, the contribution to $a_\mu$ is approximately given by
\begin{equation}
 \delta a_\mu^{\rm MGM}\simeq
  \frac{5\alpha_2+\alpha_Y}{48\pi}
\frac{m_\mu^2}{M_{\rm SUSY}^2}{\rm sign}(M_2\mu)\tan\beta.
\end{equation}
Here we assume that all the SUSY mass parameters to be equal to
$M_{\rm SUSY}$. The chargino-sneutrino loop diagram dominates over
other diagrams. The SUSY contribution is enhanced when $\tan\beta$ is
large and its sign is determined by sign($M_2 \mu$).  In the MGM, the
sign of $M_2\mu\tan\beta$ is positive, and $ \delta a_\mu^{\rm
  MGM}>0$.  This agrees with the sign of the current discrepancy of
$a_\mu$ in Eq.~(\ref{deltamu}).

\subsection{$\overline{B}\to X_s\gamma$ Decay}

The experimental world average ${\cal B}(\overline{B}\to X_s \gamma)$
by the Heavy Flavor Working Group 
\cite{Barberio:2007cr} is
\begin{equation}
 {\cal B}(\overline{B}\to X_s \gamma)_{E_{\gamma}>1.6\, {\rm GeV}} 
=(3.55 \pm 0.26) \times 10^{-4},
\end{equation}
with a photon energy cut $E_{\gamma}>1.6$ GeV.  The SM prediction of $
{\cal B}(\overline{B}\to X_s \gamma)$ has been estimated by including
NNLO contributions in Ref.~\cite{Misiak:2006zs}.  Recently, additional
perturbative corrections related with the photon energy cut has been
calculated \cite{Becher:2006pu}.  Combining those results, the most
recent theoretical prediction is given by \cite{Becher:2006pu}
\begin{equation}
 {\cal B}(\overline{B}\to X_s \gamma)_{E_\gamma >1.6\, {\rm GeV}}
 =(2.98 \pm 0.26) \times 10^{-4}.
\end{equation}
The new SM prediction becomes lower than the experimental data, and
the deviation is 1.4 $\sigma$.  While the deviation is not
significant, the recent result opens new room for additional new
physics contributions.

The $\overline{B}\to X_s\gamma$ decay is described by the following
effective Hamiltonian \cite{Misiak:2006zs},
\begin{eqnarray}
 H_{\rm eff}^{b\to s\gamma}= -\frac{4G_F}{\sqrt{2}} V_{tb}V_{ts}^\ast
 \sum_{i=1}^{8} C_i(\mu_b) O_i(\mu_b),
\end{eqnarray}
where $\mu_b\simeq m_b$.
The SUSY contributions are encoded to the Wilson coefficients, $C_i$.
The most relevant operator for the $\overline{B}\to X_s\gamma$ decay
is
\begin{eqnarray}
 O_7=\frac{e}{16\pi^2} m_b \overline{s}_L\sigma^{\mu\nu}b_R F_{\mu\nu}.
\end{eqnarray}
The calculation of  ${\cal B}(\overline{B}\to X_s \gamma)$ at NNLO
is quite complicated. Here we refer Refs.~\cite{Misiak:2006zs,Becher:2006pu}
to the formula.

In the MGM, the dominant SUSY contributions to $C_7$ come from the
charged Higgs boson, the chargino, and the gluino loop diagrams.  It is
known that the charged Higgs amplitude is always constructive with the
SM contribution, while the chargino and the gluino loop amplitudes are
either constructive or destructive with the SM one.  In the mass
insertion approximation, the pure-Higgsino $(C_7^{{\tilde h}^-})$ ,
the Higgsino-wino $(C_7^{{\tilde h}{\tilde W}^-})$, and the
gluino contributions $(C_7^{{\tilde g}})$ are written as
\cite{Rattazzi:1996fb}
\begin{eqnarray}
 C_7^{{\tilde h}^-}&\simeq&\frac{1}{2} r_b A_t \mu\tan\beta
  \frac{m_t^2}{m_{\tilde Q}^4} f_{\rm ch}(\mu^2/m_{\tilde Q}^2),
\\
C_7^{{\tilde h}{\tilde W}^-}
&\simeq&  \frac{(\delta_{LL})_{23}}{V_{tb}V^\ast_{ts}}
r_b \tan\beta\frac{M_2\mu}{M_2^2-\mu^2}
\left[f_{\rm ch}(M_2^2/m_{\tilde Q}^2)-f_{\rm ch}(\mu^2/m_{\tilde Q}^2)
\right],
\\
C_7^{\tilde g}
&\simeq& \frac{8\alpha_s}{9\alpha_2} \frac{m_W^2}{m_{\tilde Q}^2}
\frac{(\delta_{LL})_{23}}{V_{tb}V^\ast_{ts}}
r_b \tan\beta 
\frac{M_3\mu}{m_{\tilde Q}^2}
f_{\rm gl}(M_3^2/m_{\tilde Q}^2),
\end{eqnarray}
where $m_{\tilde Q}$ is an average squark mass and 
the loop functions $f_{\rm ch}$ and $f_{\rm gl}$ are defined as
\begin{eqnarray}
 f_{\rm ch}(x) &=&\frac{13-7x}{6(x-1)^3}
-\frac{3+2 x(1-x)}{3(x-1)^4}\log x,
\\
 f_{\rm gl}(x) &=&-\frac{1+10 x +x^2}{2(x-1)^4}
+\frac{3 x (1+x)}{(x-1)^5}\log x.
\end{eqnarray} 
The non-holomorphic correction to the bottom-quark Yukawa coupling is
not negligible in large $\tan\beta$ \cite{Hall:1993gn}.  The
correction is represented as $r_b= 1/(1+\epsilon_b)$, and $\epsilon_b$
in it is
\begin{eqnarray}
 \epsilon_b= \frac{2\alpha_s}{3\pi} M_3\mu\tan\beta
f\left(m^2_{\tilde Q_L},m^2_{\tilde D_R}, M_3^2
\right).
\end{eqnarray} 
The loop function $f$ in $ \epsilon_b$ is defined as
\begin{eqnarray}
 f(x,y,z)=\frac{xy \log(x/y)+yz \log(y/z)+zx \log(z/x)}
  {(x-y)(y-z)(z-x)}.
\end{eqnarray}
The flavor-violating parameter $(\delta_{LL})_{23}$ is 
induced by the RGE effect and is written in the leading logarithmic
approximation as 
\begin{eqnarray}
(\delta_{LL})_{23}\equiv\frac{(m^2_{\tilde Q})_{23}}{m^2_{\tilde Q}}
\simeq -\frac{\lambda_t^2}{4\pi^2} V_{tb}V_{ts}^\ast 
\log\left(\frac{M_M}{M_3}\right).
\end{eqnarray}

With the boundary condition $\overline{A}=0$, $A_t$ is induce by the
RGE effect and proportional to the gaugino masses.  As discussed in
the previous section, sign($M_2\mu\tan\beta$) is predicted to be
positive in the MGM.  In this case sign($A_t\mu\tan\beta$) is also
positive and the sign of the pure-Higgsino amplitude is opposite to
that of the charged Higgs contribution.  Then the dominant SUSY
contributions can cancel with each other. The Higgsino-wino and gluino
contributions are also constructive with the pure-Higgsino one. 

In order to calculate ${\cal B}(\overline{B}\to X_s \gamma)$ at NNLO
in the MGM, SUSY contributions should be matched into $C_i$ beyond LO
at the weak scale.  But this is beyond the scope of this paper.  For
the numerical calculation, we use the LO SUSY corrections to $C_i$ at
the weak scale while we have included the SM contributions at NNLO.

\subsection{${\overline B}\to X_s l^+ l^-$ Decay}

${\cal B}(\overline{B} \to X_s l^+ l^-)$ in a low dilepton invariant
mass region, $1<m_{ll}^2<6\, {\rm GeV}^2$, was reported by Babar,
$(1.8\pm 0.9)\times 10^{-6}$ \cite{Aubert:2004it} and Belle, $(1.5\pm
0.6)\times 10^{-6}$ \cite{Iwasaki:2005sy}.  The current world
average is given by
\begin{eqnarray}
 {\cal B}(\overline{B} \to X_s l^+ l^-)_{1<m_{ll}^2<6\, {\rm GeV}^2}
 = (1.6\pm 0.51)\times 10^{-6}.
\end{eqnarray}
The branching ratio in a higher dilepton invariant mass region is also
measured.  In a higher $m_{ll}$ region the branching ratio suffers
from long-distance contributions of $J/\psi$ and $\psi$' resonances,
and a large theoretical uncertainty is expected. However, in
$0<m_{ll}^2<6$ GeV$^2$ region, the short-distance contribution
dominates, and the branching ratio is sensitive to the new physics
contribution. Therefore in this paper, we consider $ {\cal
  B}(\overline{B} \to X_s l^+ l^-)$ in the low dilepton region.  The
SM prediction of $ {\cal B}(\overline{B} \to X_s \mu^+ \mu^-)$ at NNLO
of QCD including a QED correction is given by \cite{Huber:2005ig}
\begin{eqnarray}
 {\cal B}(\overline{B} \to X_s \mu^+ \mu^-)_{1<m_{ll}^2<6\, {\rm GeV}^2}
 = (1.59\pm 0.11)\times 10^{-6}.
\end{eqnarray}
The SM prediction is consistent with the current experimental data.

The $\overline{B} \to X_s l^+ l^-$ decay is described by the
effective Hamiltonian
\begin{eqnarray}
 H_{\rm eff}^{b\to sl^+l^-}=H_{\rm eff}^{b\to s\gamma}
-\frac{4G_F}{\sqrt{2}} V_{tb}V_{ts}^\ast
\left[C_9(\mu_b) O_9(\mu_b)
+C_{10}(\mu_b) O_{10}(\mu_b)\right],
\end{eqnarray}
where new operators $O_9$ and  $O_{10}$ are defined as
\begin{eqnarray}
 O_9&=&  (\overline{s_L}\gamma_\mu
  b_L)(\overline{l}\gamma^\mu l),
\\
 O_{10}&=&  (\overline{s_L}\gamma_\mu
  b_L)(\overline{l}\gamma^\mu \gamma_5l).
\end{eqnarray}
In the MGM, the Wilson coefficients $C_9$ and $C_{10}$ are dominated
by the SM contributions, and the SUSY particles can contribute to
${\cal B}(\overline{B} \to X_s l^+ l^-)$ mainly through $C_7$. As a
result, the SUSY contributions to ${\cal B}(\overline{B} \to X_s
\gamma)$ and ${\cal B}(\overline{B} \to X_s \mu^+ \mu^-)$ are strongly
correlated.

The branching ratio in the low dilepton mass region is written by 
\cite{Huber:2005ig}
\begin{eqnarray}
 {\cal B}(\overline{B} \to X_s \mu^+ \mu^-)_{1<m_{ll}^2<6\, {\rm GeV}^2}
&=&
 (2.19-0.543 R_{10}+0.0281 R_7+0.0153 R_7 R_{10} 
\nonumber\\
&+& 0.0686 R_7 R_8 -0.870 R_7 R_9-0.0128 R_8 +0.00195 R_8 R_{10}
\nonumber\\
&-&0.0993 R_8 R_9+2.84 R_9
-0.107 R_9 R_{10} + 11.0 R_{10}^2
\nonumber\\
&+&0.281 R_7^2+0.00377 R_8^2  + 1.53
R_9^2)\times 10^{-7}.
\end{eqnarray}
Here $R_i$'s are ratios of the total and the SM contributions,
\begin{eqnarray}
 R_{7,8}=\frac{C_{7,8}^{{\rm eff}}(\mu_W)}
          {C_{7,8}^{{\rm eff,SM}}(\mu_W)},~~~~~
 R_{9,10}=\frac{C_{9,10}(\mu_W)}
          {C_{9,10}^{{SM}}(\mu_W)},
\end{eqnarray}
where $C_{i}^{\rm eff}$'s are the effective Wilson coefficients at
$\mu_W\simeq m_t$ and they are the LO contributions
\cite{Huber:2005ig}. 

${\cal B}(\overline{B} \to X_s \gamma)$ depends mainly on $|C_7|^2$,
and it is insensitive to the sign of $C_7$.  However, ${\cal
  B}(\overline{B} \to X_s l^+ l^-)$ depends on the sign of $C_7$ since
there is an interference term between $R_7$ and $R_9$.  When the sign
of $C_7$ is opposite to the SM prediction, ${\cal B}(\overline{B} \to
X_s l^+ l^-)$ becomes too large compared with the current experimental
data, and such a possibility is strongly disfavored
\cite{Gambino:2004mv}.

\subsection{$B_s\to \mu^+\mu^-$ Decay}

The current upper bound on ${\cal B}(B_s\to\mu^+\mu^-)$ is 
given by the CDF collaboration \cite{Scuri:2007py},
\begin{equation}
 {\cal B}(B_s\to\mu^+\mu^-) < 1\times 10^{-7},
\end{equation}
at 95\% C.L. In the SM  $ {\cal B}(B_s\to\mu^+\mu^-)$ is
suppressed by the muon mass and quite small. The SM prediction
is estimated as ${\cal B}(B_s\to\mu^+\mu^-)_{\rm SM}=(3.35\pm 0.32)
\times 10^{-9}$ \cite{Blanke:2006ig}. The current experimental upper bound is
about two orders of magnitude larger than the SM prediction, and
a clue to new physics may be hidden there.

The branching ratio of $B_s\to \mu^+\mu^-$ is written by 
\cite{Bobeth:2001sq}
\begin{eqnarray}
 {\cal B}(B_s\to\mu^+\mu^-) 
&&=\frac{G_F^2\alpha_2^2 \tau_B m_B^5 f^2_{B}}{64\pi^3}
\left|V_{tb}V_{ts}\right|^2
\sqrt{1-\frac{4 m_\mu^2}{m_B^2}}
\nonumber\\
&&\times\left\{\left(1-\frac{4 m_\mu^2}{m_B^2}\right)
\left|\frac{C_S-C_S^{'}}{m_b+m_s}\right|^2
+\left|\frac{C_P-C_P^{'}}{m_b+m_s}+2\frac{m_\mu}{m_B^2}(C_A-C_A^{'})\right|
\right\}.
\end{eqnarray}
Here $C_s^{(')}$, $C_P^{(')}$, and $C_A^{(')}$ are the Wilson
coefficients which include the SUSY contributions.  In
Ref.~\cite{Babu:1999hn}, it was pointed out that the amplitude of
neutral Higgs boson exchange diagram is proportional to $\tan^3\beta$
in a large $\tan\beta$ region due to non-holomorphic correction to
the Yukawa coupling.  The dominant SUSY contribution is written by
\begin{eqnarray}
 C_S\simeq -C_P\simeq 
\frac{m_b m_\mu m_t^2}{4 m_W^2 m_A^2}r_b^2 \tan^3\beta A_t \mu 
f\left(m^2_{\tilde Q_L},m^2_{\tilde U_R}, \mu^2
\right),
\end{eqnarray}
and $C_{S}^{'}=(m_s/m_b)C_{S}$ and $C_{P}^{'}=-(m_s/m_b)C_{P}$.  The
dominant SUSY contribution to ${\cal B}(B_s\to\mu^+\mu^-)$ is
proportional to $\tan^6\beta$ and can be enhanced by orders of
magnitude when $\tan\beta\gsim 10$.  Since $r_b<1$ in the MGM, the
$\tan\beta$ enhancement becomes mild.  For the numerical calculation
we use the complete expression in
Refs.~\cite{Isidori:2001fv,Buras:2002wq}.

\subsection{$B^-\to\tau^-\,\overline{\nu}$  Decay}

The Belle Collaboration reported evidence of $B^-\to\tau^-\,\overline{\nu}$, 
and the measured branching ratio is  
\cite{Ikado:2006un}
\begin{equation}
 {\cal B}(B^- \to \tau^- \,\overline{\nu}) 
=(1.79^{+0.56}_{-0.49}(stat.)^{+0.46}_{-0.51}(syst.))\times {10^{-4}}.
\end{equation}
In the SM the branching ratio is given by 
\begin{equation}
 {\cal B}(B^-\to\tau^-\,\overline{\nu})=\frac{G_F^2m_Bm_\tau^2}{8\pi}
\left(1-\frac{m_\tau^2}{m_B^2}\right)\tau_{B^-}f_B^2 \left|V_{ub}\right|^2.
\end{equation}
Using $|V_{ub}|=(4.31\pm 0.30)\times 10^{-3}$ \cite{Yao:2006px} 
and  $f_B=0.216\pm 0.022$ GeV \cite{Gray:2005ad}, 
the SM prediction is ${\cal B}(B^-\to\tau^-\,\overline{\nu})=
(1.6\pm 0.4)\times 10^{-4}$. 
The current experimental data is consistent with the SM prediction.

In the MGM, the charged Higgs boson exchange contributes to 
the $B^-\to\tau^-\,\overline{\nu}$ decay. The branching ratio is written by
\begin{equation}
 {\cal B}(B^-\to\tau^-\,\overline{\nu})=\frac{G_F^2m_Bm_\tau^2}{8\pi}
\left(1-\frac{m_\tau^2}{m_B^2}\right)\tau_{B^-}f_B^2
\left|V_{ub}\right|^2 
\left(1- r_b\tan^2\beta\frac{m^2_B}{m_{H^-}^2}\right)^2,
\end{equation}
where the non-holomorphic correction $r_b$ is included \cite{Akeroyd:2003zr}.
The charged Higgs boson amplitude interferes destructively with 
SM contribution and become significant when $\tan\beta$ is large.
The current experimental data gives the constraint on the charged Higgs 
contribution. However, a large $\tan\beta$ region is still allowed
since the cancellation occurs between the SM and the charged 
Higgs contributions.

\section{Numerical Analysis}

\begin{figure}
 \centerline{
\epsfxsize = 0.5\textwidth \rotatebox{-90}{\epsffile{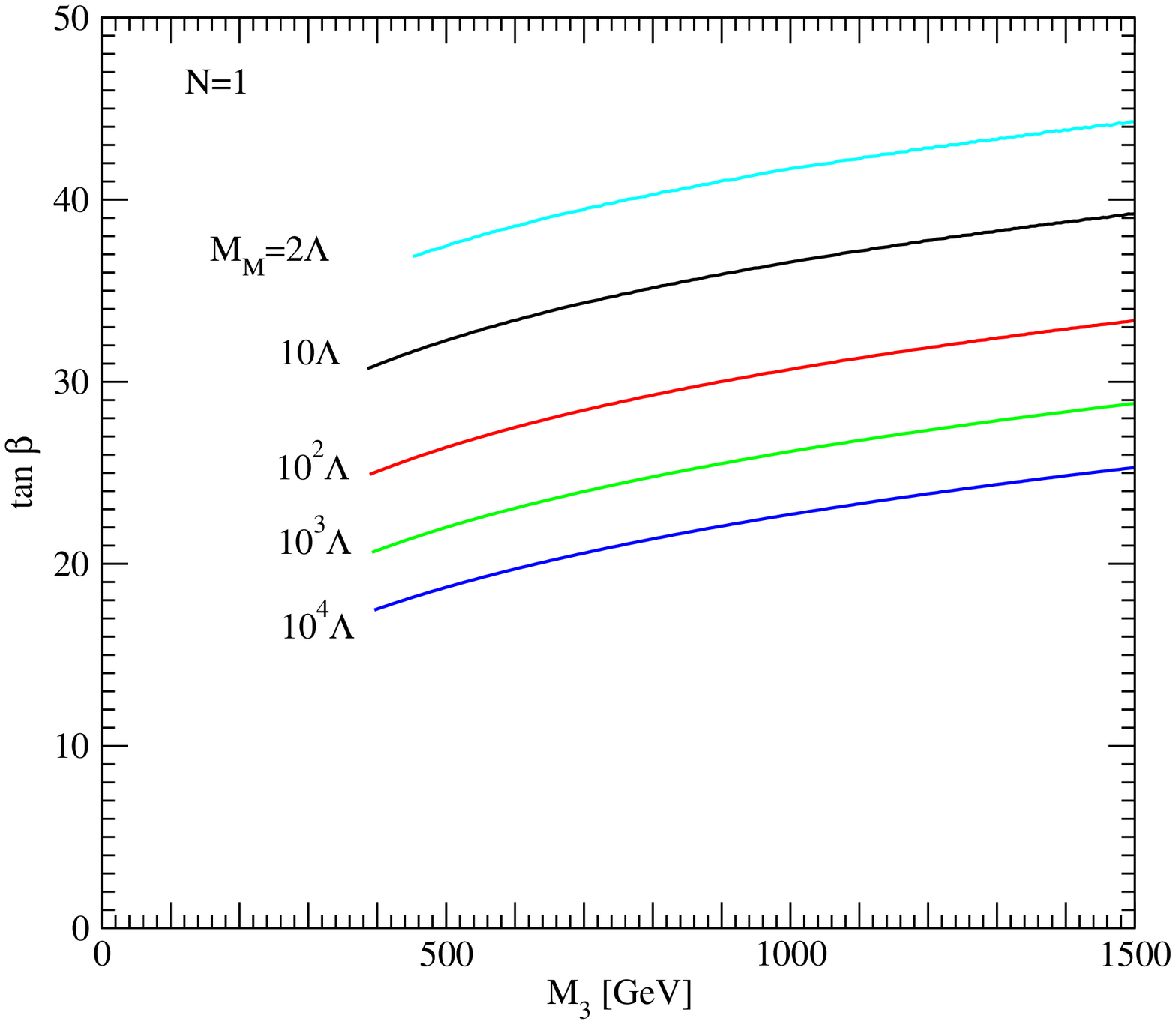} }
\epsfxsize = 0.5\textwidth \rotatebox{-90}{\epsffile{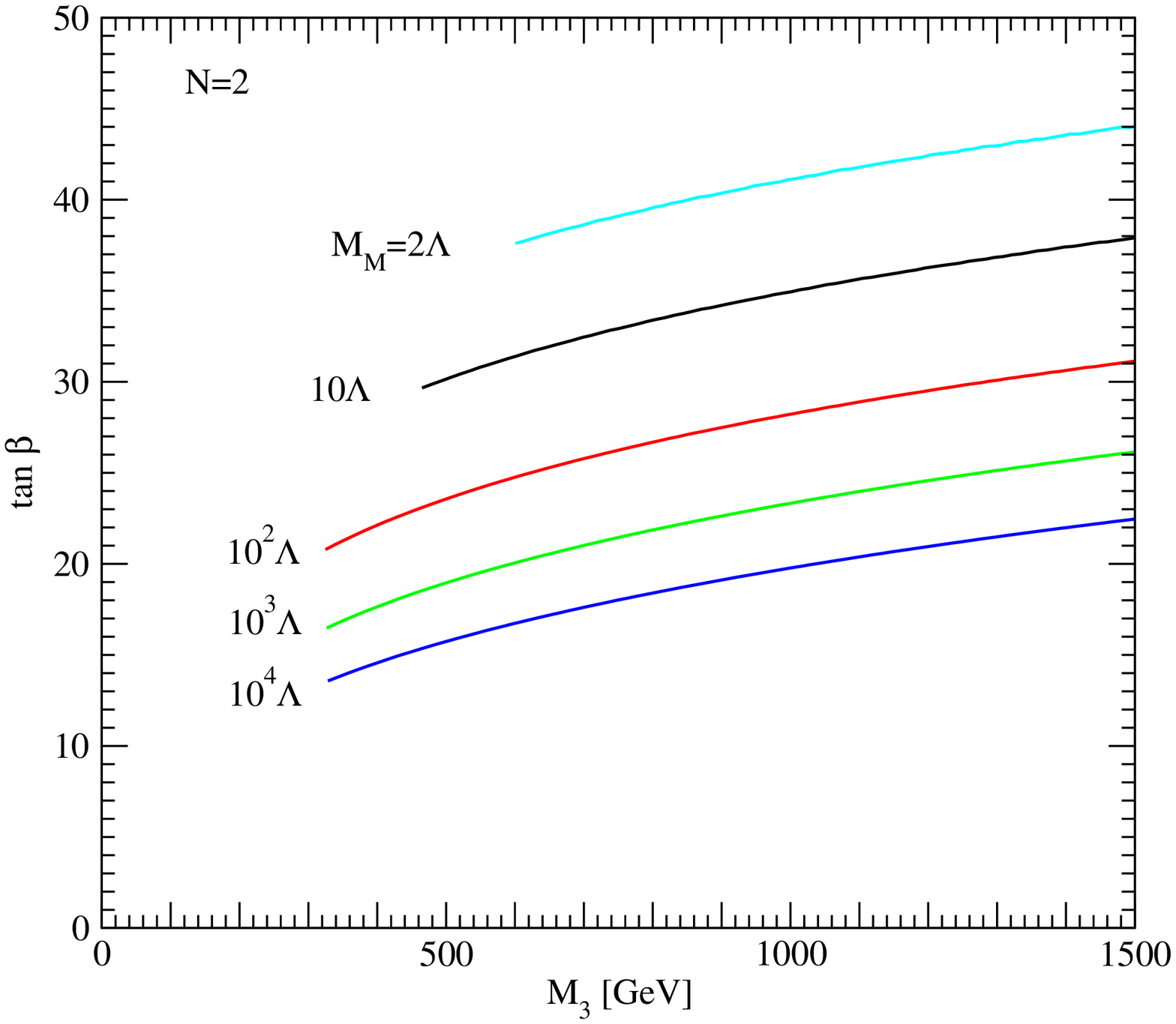} }
}
 \centerline{
\epsfxsize = 0.5\textwidth \rotatebox{-90}{\epsffile{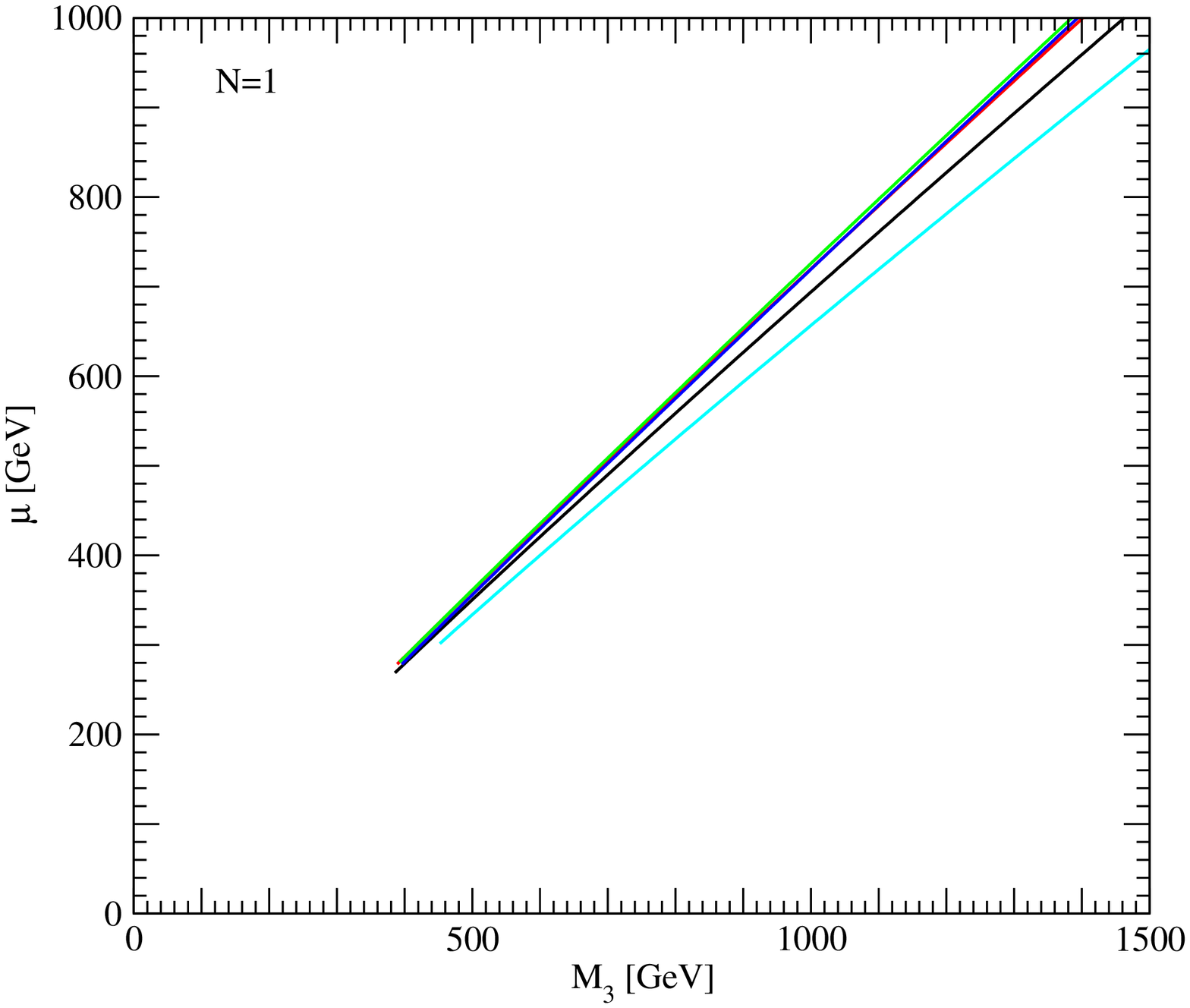} }
\epsfxsize = 0.5\textwidth \rotatebox{-90}{\epsffile{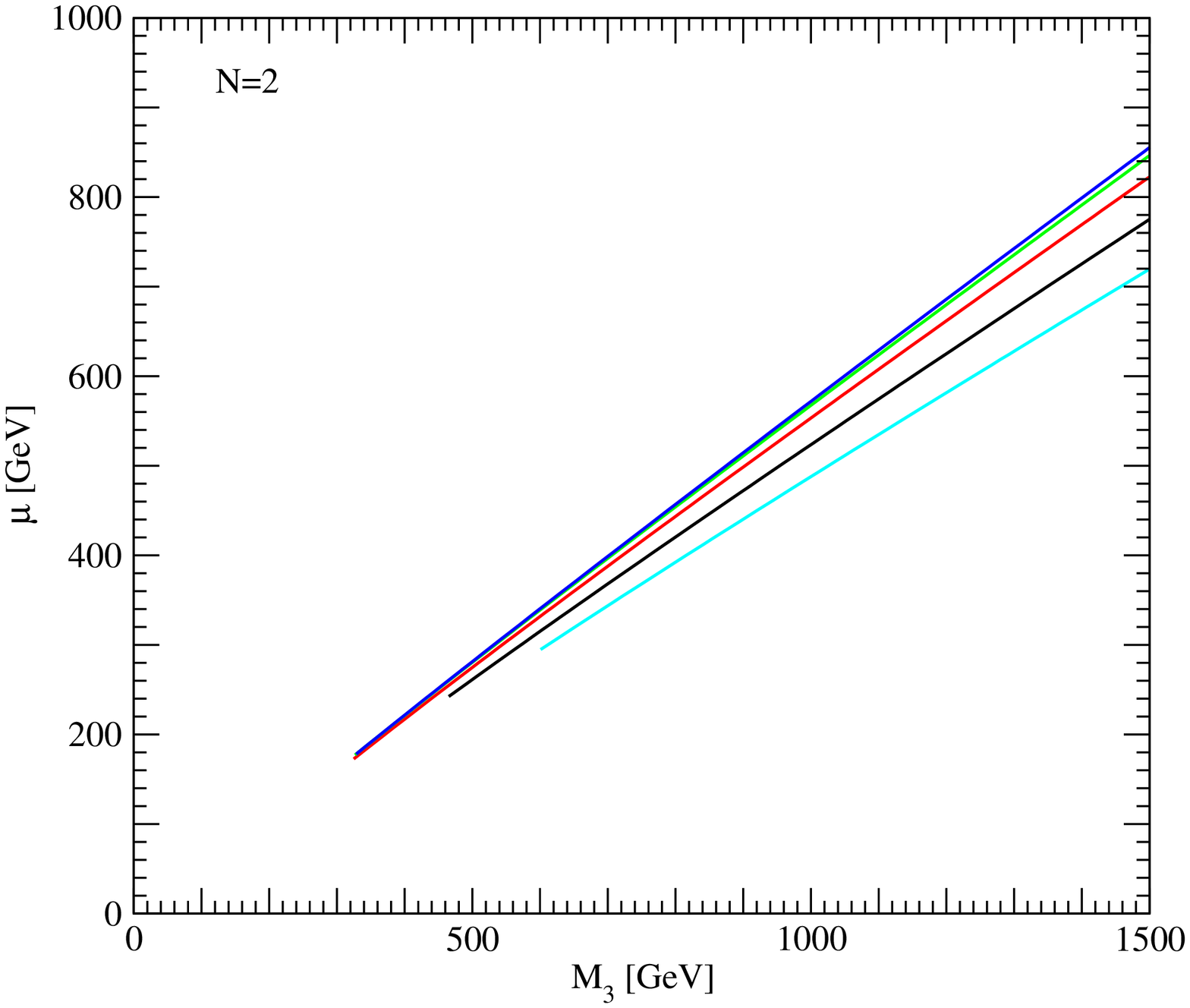} }
}
\vspace*{-5mm}
\caption{
Two parameters, $\mu$ and $\tan\beta$, as functions of $M_3$
  for $N=1$ (left) and $2$ (right). Each line in figures of $\mu$
  parameter is for $M_M=2,10,10^2,10^3$, and $10^4\Lambda$ from the
  bottom line.  
}
\label{figmutanb}
\end{figure}

In the MGM, there are three input parameters, $\Lambda$, $M_M$, and $N$.
The numerical results do not depend strongly on the difference between
$M_{M_3}$ and $M_{M_2}$, so we assume that $M_{M_3}=1.3 M_{M_2}(\equiv M_M)$ in
the following analysis.
The gaugino masses are almost determined by $\Lambda$ so we will show
the numerical result as a function of the gluino mass $M_3$.  We
consider the messenger scale as $M_{M}/\Lambda =2,10,10^2,10^3$, and
$10^4$.  For the quark masses, we use $m_t=172.5$ GeV and $m_b=4.7$
GeV \cite{Yao:2006px}.
  With the inputs at the messenger scale, we can calculate the
SUSY breaking parameters at the weak scale by solving the RGE of the
MSSM.  As discussed in Section~\ref{MGM}, $\tan\beta$ and $\mu$ are
determined from the input parameters by imposing the correct
electroweak symmetry breaking. Once all the SUSY breaking parameters
at the weak scale are determined, the SUSY contributions to the
low-energy observables can be evaluated.
For convenience, $\tan\beta$ and $\mu$ are shown in Fig.~\ref{figmutanb}
as functions of $M_3$ for various $M_M$'s.

\begin{figure}
 \centerline{
\epsfxsize = 0.5\textwidth \rotatebox{-90}{\epsffile{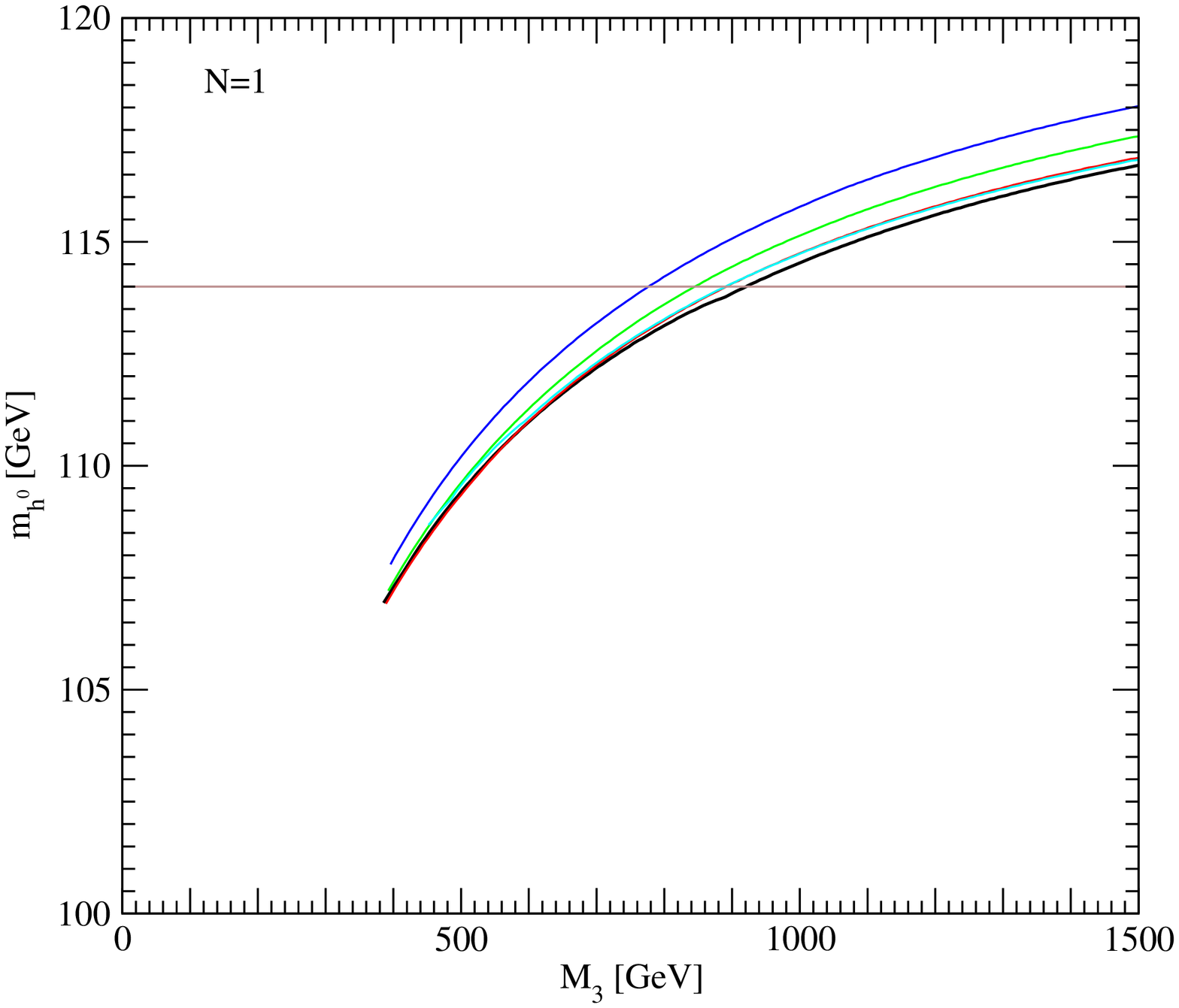} }
\epsfxsize = 0.5\textwidth \rotatebox{-90}{\epsffile{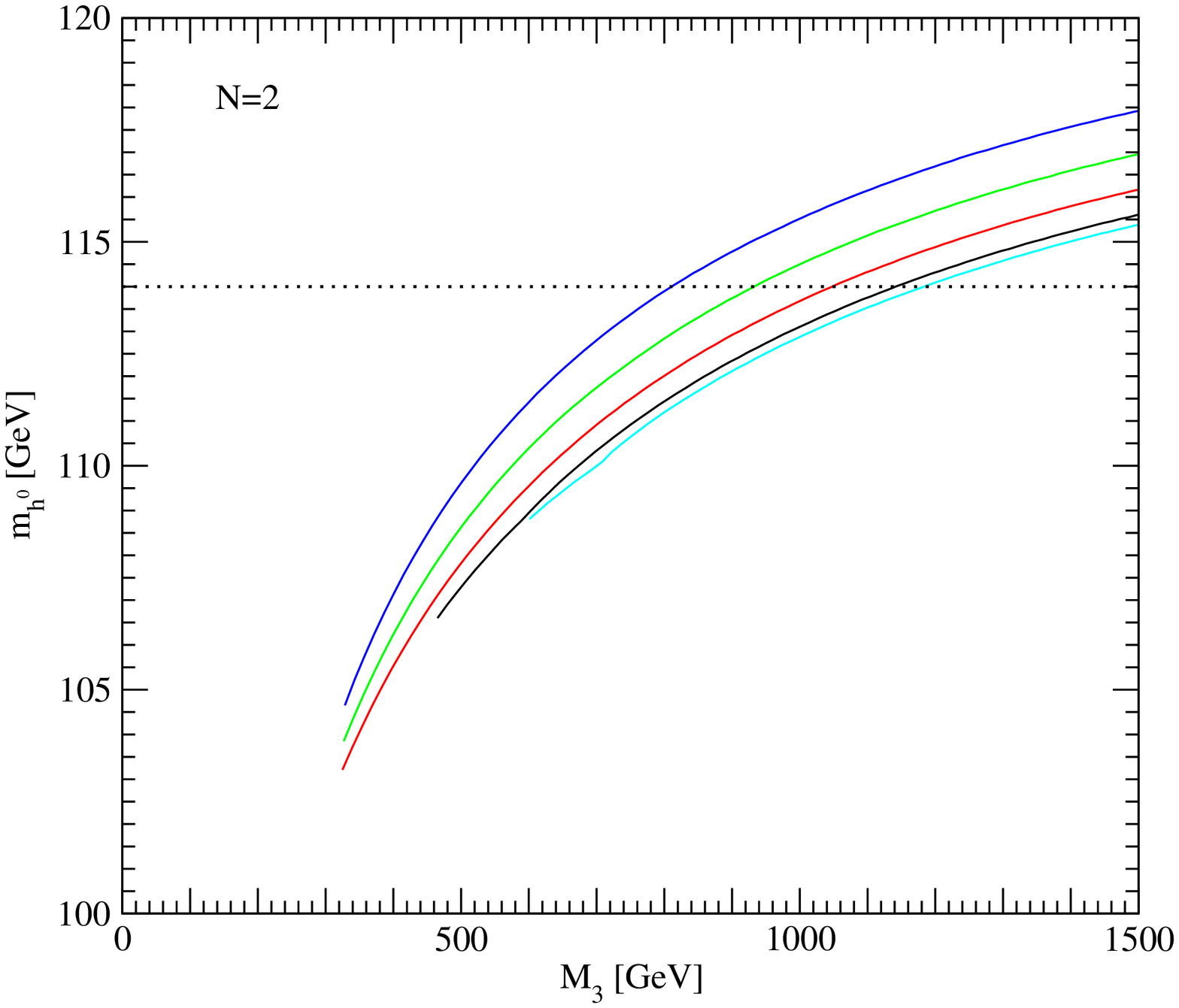} }
}
\vspace*{-5mm}
\caption{ Light neutral Higgs boson mass as a function of $M_3$ for
  $N=1$ (left) and $2$ (right). Each line is for $M_M=2,10,10^2,10^3$,
  and $10^4\Lambda$ from the bottom line.  }
\label{figmh}
\end{figure}

First, we discuss the light Higgs boson boson mass, which is shown in
Fig.~\ref{figmh} as a function of $M_3$ for various $M_M$'s.  We have
calculated the Higgs mass using the FeynHiggs code
\cite{Hahn:2005cu}. The Higgs boson mass depends mainly on $M_3$ ({\it
  i.e. }$\Lambda$).  From the LEP bound on the neutral Higgs boson
mass $(m_h>114.4~{\rm GeV})$, we can obtain the following lowerbound
on the $M_3$,
\begin{equation}
 M_3 \gsim 800\sim 900~ (800\sim1100)~{\rm GeV},  ~~~{\rm for }~N=1\,(2).
\end{equation}
We find that this gives strong bounds on signatures of the MGM.

It is pointed out in Ref.~\cite{Rattazzi:1996fb} that the stability of
the charge-conserving vacuum gives a lowerbound on $M_3$ in the MGM.
When $\tan\beta$ is quite large, the lightest stau becomes light.
In addition, the trilinear coupling of the left- and right-handed
staus with the Higgs field, which is enhanced by $\tan\beta$, may
destabilize the charge-conserving vacuum. However, it is found from
Fig.~1 in Ref.~\cite{Rattazzi:1996fb} that the lowerbound on $M_3$ is
about $700\sim 800$ GeV even for $M_M\simeq 2\Lambda$ when $N=1$, and
larger $M_M$ makes the bound weaker. When increasing $N$, the
lowerbound is naively expected to be scaled by $\sqrt{N}$.  Thus, the
bound on $M_3$ from the Higgs boson mass is comparable to or 
 rather stronger than that of the vacuum stability.

\begin{figure}
 \centerline{
\epsfxsize = 0.5\textwidth \rotatebox{-90}{\epsffile{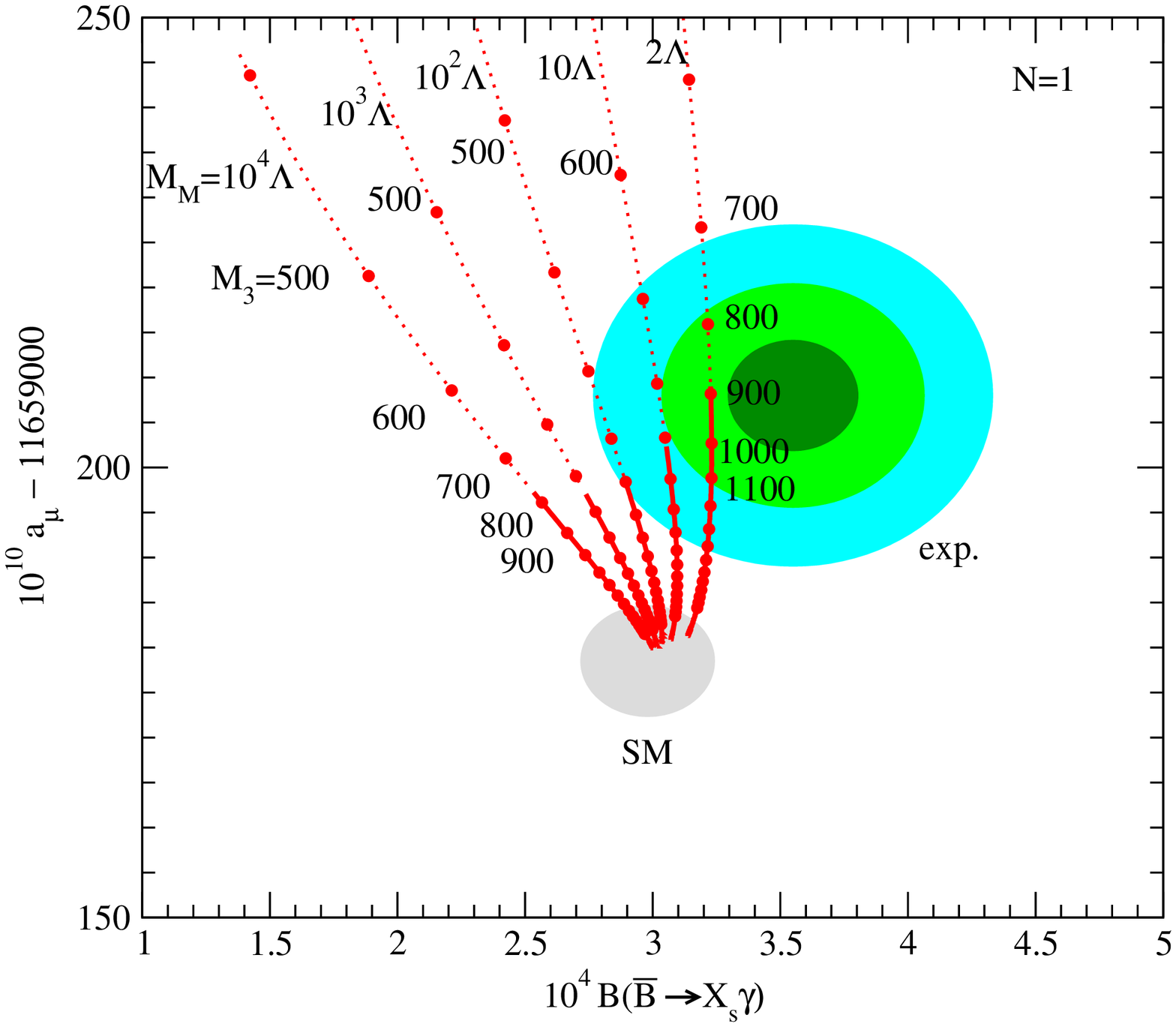} }
\epsfxsize = 0.5\textwidth \rotatebox{-90}{\epsffile{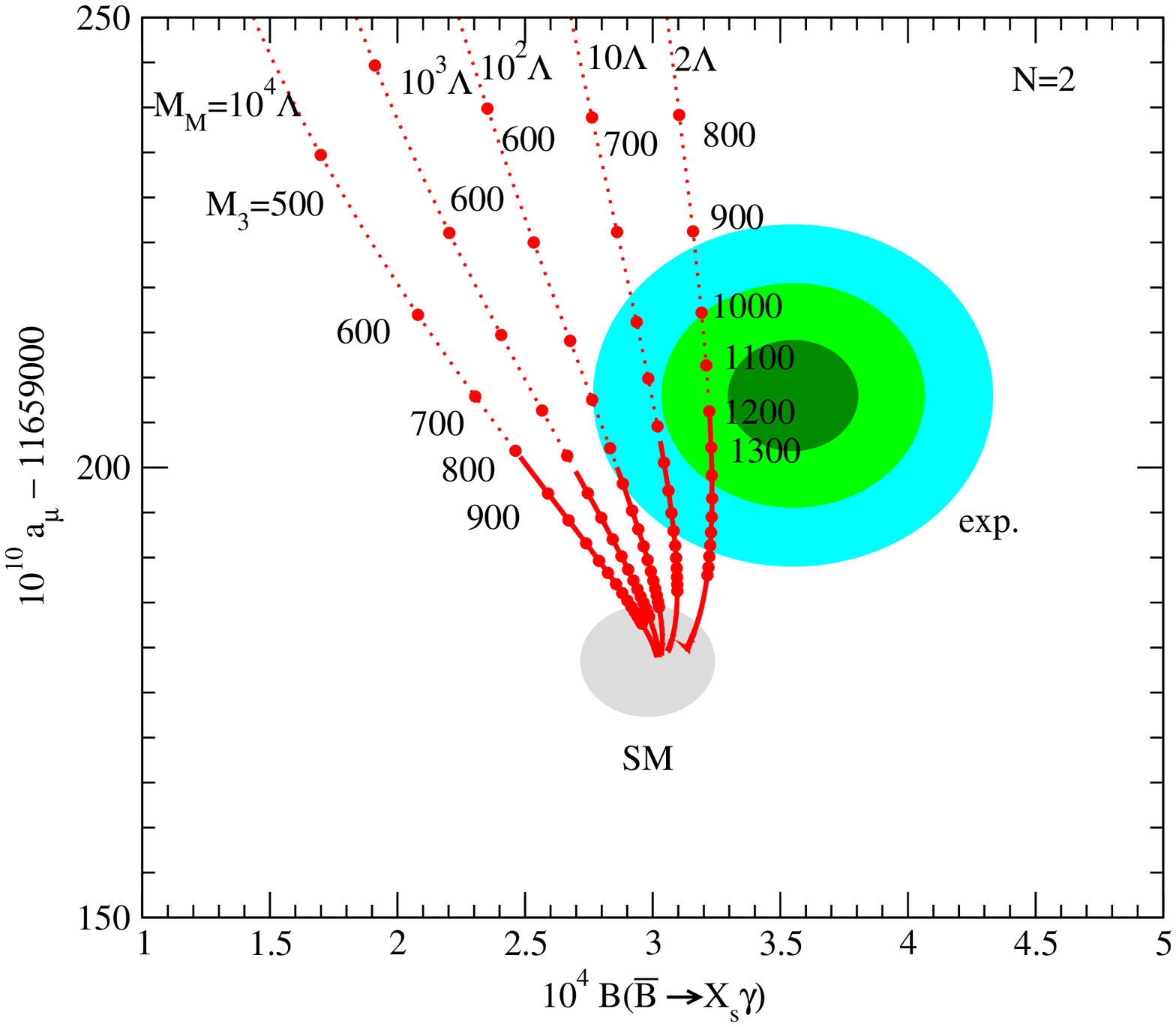} }
}
\vspace*{-5mm}
\caption{
Correlation between ${\cal B}(\overline{B}\to X_s \gamma)$ and
$a_\mu$ in the MGM for $N=1$ and $2$. The colored shaded regions
represent experimentally allowed ones at 1,2, and 3 $\sigma$,
respectively. The grey shaded region is theoretical prediction
with 1 $\sigma$ uncertainty. The solid (dotted) lines 
are experimentally allowed (excluded) regions.
The circle dots represent the discrete $M_3$'s with 100 GeV intervals.
}
\label{bsg-amu}
\end{figure}

Now we discuss the correlation between ${\cal B}(\overline{B}\to X_s
\gamma)$ and $a_\mu$ for various $M_M$. It is shown in
Fig.~\ref{bsg-amu}. The left (right) figure is for $N=1\,(2)$.  In
these figures, the colored shaded regions are the experimental bounds
at 1, 2, and 3 $\sigma$ levels, and the dark shaded region is the SM
model prediction with a 1 $\sigma$ error.  The red circle dots
correspond to discrete $M_3$'s with 100 GeV intervals.  The allowed
regions are plotted with solid lines while the regions experimentally
excluded by the light Higgs mass bound are also plotted with dotted
lines.

 We can see that the SM predictions for these processes are rather low
compared to the experimental data. The sizes of the experimental
errors are almost comparable to those for theoretical ones. Now we stand 
on a stage in which we can pin down a parameter region in models beyond 
the SM by combining  experimental results of  low-energy observables. 

As discussed in Section~\ref{amu}, the SUSY contribution to $a_\mu$
are always positive. The magnitude of the SUSY contribution is mainly
determined by $M_3(\Lambda)$ and is not so sensitive to $M_M$ and $N$.
On the other hand, the SUSY contributions to ${\cal B}(\overline{B}\to
X_s \gamma)$ depends on all the parameters.  

For smaller $M_M$ ${\cal B}(\overline{B}\to X_s \gamma)$ is larger
than the SM prediction, while ${\cal B}(\overline{B}\to X_s \gamma)$
can be smaller than the SM prediction in relatively small $M_3$
region.  We find that the chargino and the gluino contribution are not
significant in small $M_M$ regions, in which the non-SM contributions
to $C_7$ are dominated by the charged Higgs contribution.  Since the
charged Higgs amplitude always contributes to $C_7$ constructively,
${\cal B}(\overline{B}\to X_s \gamma)$ in the MGM becomes larger in
small $M_M$ region.  For large $M_M$ region, the chargino contribution
becomes significant and dominates the SUSY contribution to $C_7$. As a
result, ${\cal B}(\overline{B}\to X_s \gamma)$ can be smaller than the
SM prediction.  

We find that the most favorable region in the MGM from ${\cal
  B}(\overline{B}\to X_s \gamma)$ and $a_\mu$ is $M_M\simeq 2\Lambda$
and $M_3\simeq 900~(1200)$ GeV for $N=1\, (2)$. The gluino mass will be
covered by the direct SUSY search in the LHC experiment.  The light
Higgs boson mass is close to the current experimental lowerbound as in
Fig.~\ref{figmh}.

\begin{figure}

 \centerline{
\epsfxsize = 0.5\textwidth \rotatebox{-90}{\epsffile{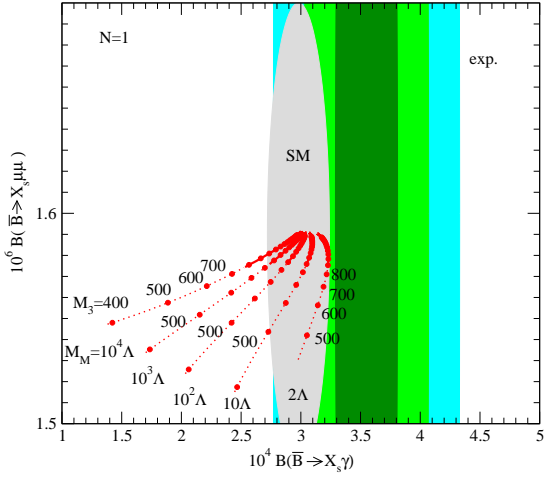}}
\hspace{.5cm}
\epsfxsize = 0.5\textwidth \rotatebox{-90}{\epsffile{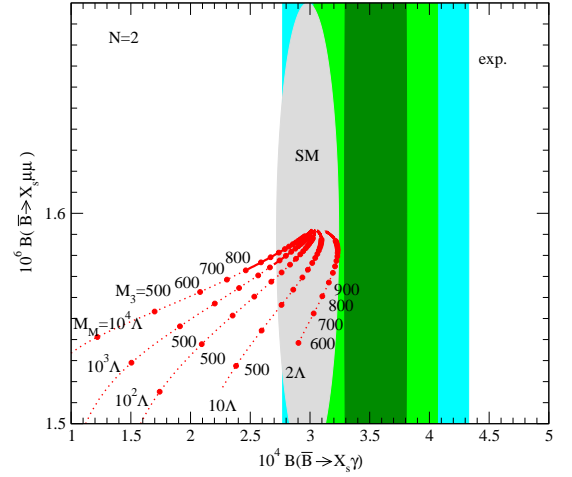} }
}
\vspace*{-5mm}
\caption{
Correlation between ${\cal B}(\overline{B}\to X_s \gamma)$ and
${\cal B}(\overline{B} \to X_s l^+ l^-)$ in the MGM for $N=1$ and $2$.
}
\label{bsg-ll}
\end{figure}

In Fig.~\ref{bsg-ll}, we show the correlation between ${\cal
  B}(\overline{B}\to X_s \gamma)$ and ${\cal B}(\overline{B}\to X_s
\mu^+\mu^-)$.  The SUSY contributions to ${\cal B}(\overline{B}\to X_s
\mu^+\mu^-)$ is negative and rather small.  If the constraint from the
light Higgs boson mass is taking into account, the deviation is at
most few \%.  Considering both theoretical and experimental
uncertainties, it would be very difficult to observe the deviation of
${\cal B}(\overline{B}\to X_s \mu^+\mu^-)$ in near future.

\begin{figure}
 \centerline{
\epsfxsize = 0.5\textwidth \rotatebox{-90}{\epsffile{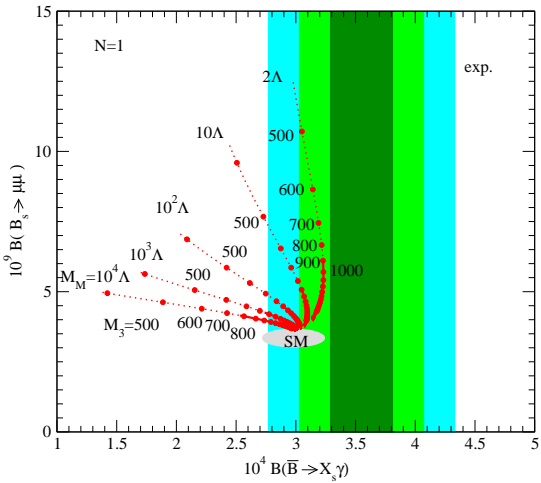}}
\hspace{.5cm}
\epsfxsize = 0.5\textwidth \rotatebox{-90}{\epsffile{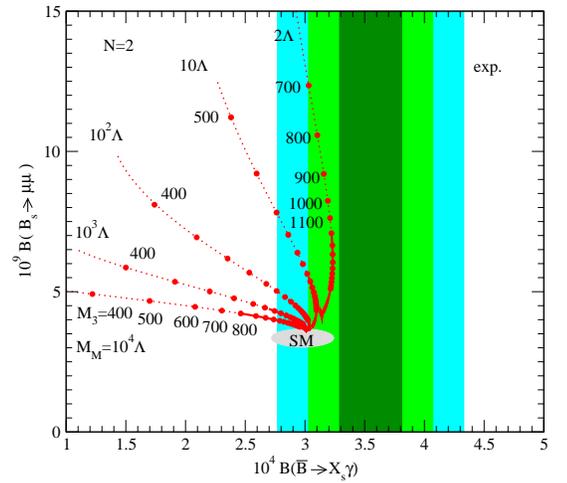} }
}
\vspace*{-5mm}
\caption{
Correlation between ${\cal B}(\overline{B}\to X_s \gamma)$ and
$ {\cal B}(B_s\to\mu^+\mu^-) $ in the MGM for $N=1$ and $2$.
}
\label{bsg-mm}
\end{figure}

In Fig.~\ref{bsg-mm}, we show the correlation between ${\cal
  B}(\overline{B}\to X_s \gamma)$ and ${\cal B}(B_s\to \mu^+\mu^-)$.
${\cal B}(B_s\to \mu^+\mu^-)$ becomes large for small $M_M$ regions,
since $\tan\beta$ becomes very large $(\simeq 40-50)$, and the
$\tan^6\beta$ enhancement on ${\cal B}(B_s\to \mu^+\mu^-)$ is
effective.  When we consider the favorable region from ${\cal
  B}(\overline{B}\to X_s \gamma)$ and $a_\mu$, the branching ratio is
estimated as ${\cal B}(B_s\to \mu^+\mu^-)\simeq (6-7)\times 10^{-9}$.
This is still far below the current experimental bound, but with in a
reach of search at LHC.

\begin{figure}
 \centerline{
\epsfxsize = 0.5\textwidth \rotatebox{-90}{\epsffile{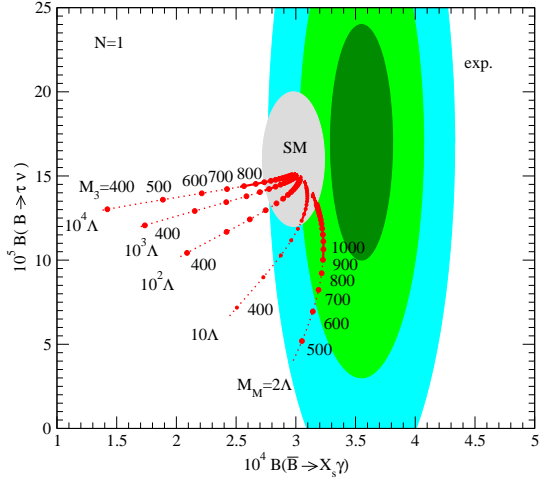}}
\hspace{.5cm}
\epsfxsize = 0.5\textwidth \rotatebox{-90}{\epsffile{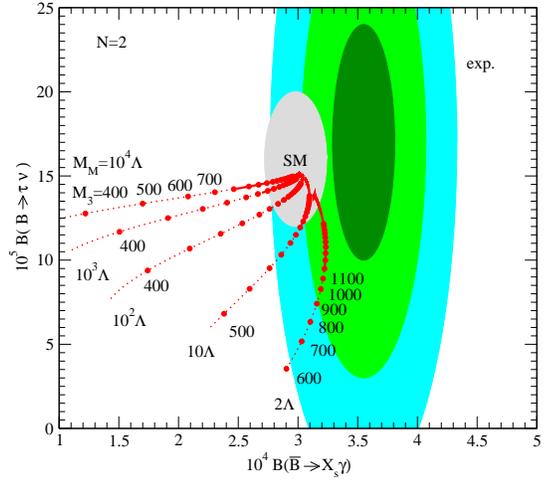} }
}
\vspace*{-5mm}
\caption{
Correlation between ${\cal B}(\overline{B}\to X_s \gamma)$ and
$ {\cal B}(B^-\to\tau^-\,\overline{\nu}) $ in the MGM for $N=1$ and $2$.
}
\label{bsg-nu}
\end{figure}

Finally, in Fig.~\ref{bsg-nu}, we show the correlation between ${\cal
  B}(\overline{B}\to X_s \gamma)$ and ${\cal B}(B^-\to
\tau^-\overline{\nu})$. The deviation of ${\cal B}(B^-\to
\tau^-\overline{\nu})$ is always negative in the MGM and become
significant for small $M_M$ regions. This is because $\tan\beta$
becomes very large in small $M_M$ and the charged Higgs contribution
interferes destructively with the SM contribution.  At the favorable
region from ${\cal B}(\overline{B}\to X_s \gamma)$ and $a_\mu$, ${\cal
  B}(B^-\to \tau^-\,\overline{\nu}) \simeq 1 \times 10^{-4}$. Notice
that this branching ratio is around the 1 $\sigma$ lower bound of
current experimental data.  A precious measurement of ${\cal B}(B^-\to
\tau^-\,\overline{\nu})$ at $B$ factories may be helpful for search for
signature of the MGM.

\begin{figure}
 \centerline{
\epsfxsize = 0.5\textwidth \rotatebox{-90}{\epsffile{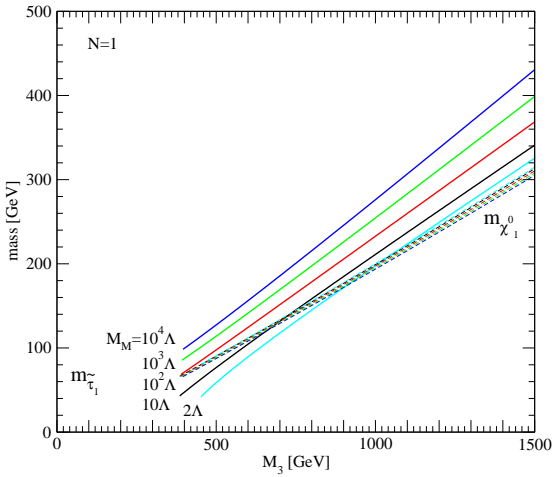}}
\hspace{.5cm}
\epsfxsize = 0.5\textwidth \rotatebox{-90}{\epsffile{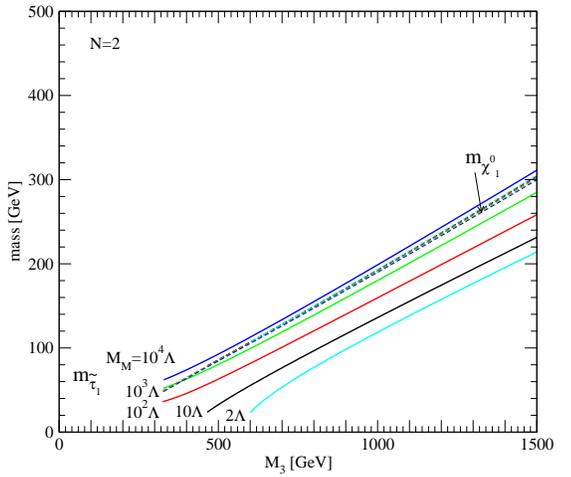} }
}
\vspace*{-5mm}
\caption{
Lightest stau and lightest neutralino masses as functions of
  $M_3$ for $N=1$ (left) and $2$ (right).  
}
\label{mstau-chi0}
\end{figure}

\section{Summary and Discussion}

We have revisited phenomenology in the minimal gauge mediated model,
in which $\tan\beta$ is naturally large.  We have considered the
anomalous magnetic moment of the muon, the branching ratios of
$\overline{B}\to X_s \gamma$, $\overline{B}\to X_s l^+l^-$, $B_s\to
\mu^+\mu^-$ and $B^-\to \tau^-\,\overline{\nu}$.  When $\tan\beta$ is
very large, the SUSY contributions to those observables can be
enhanced, and the deviations from the SM predictions are strongly
correlated with each other. We have updated the results in
\cite{Gabrielli:1998sw} taking account of the recent theoretical and
experimental developments of $\overline{B}\to X_s \gamma$ and the
anomalous magnetic moment of the muon.

We have shown that the experimental bound on the light Higgs boson gives 
a strong bound on the MGM and requires relatively heavy SUSY spectrum.
We find that the lower bound of the gluino mass is  
$M_3 \gsim 800 \sim 900$ $(800\sim 1100)$ GeV for $N=1\,(2)$. 
This constraint makes the signatures in the MGM less significant. 

When the messenger scale is larger than $\Lambda$ by orders of magnitude,
${\cal B}(\overline{B}\to X_s \gamma)$ is smaller than the SM prediction,
since the chargino loop contribution interferes destructively with the
SM one. On the other hand, when the messenger scale is same order of
$\Lambda$, ${\cal B}(\overline{B}\to X_s \gamma)$ becomes larger than the SM
prediction since the charged Higgs contribution becomes important.

The most recent theoretical calculation of ${\cal B}(\overline{B}\to
X_s \gamma)$ in the SM \cite{Misiak:2006zs, Becher:2006pu} is
1.4~$\sigma$ lower than the experimental world average
\cite{Barberio:2007cr}. When combining it with a 3.4 $\sigma$ 
deviation in $a_\mu$, a parameter region in $M_M\sim \Lambda$ and
$M_3\sim 1$~TeV is favored. In the region, ${\cal
  B}(B^-\to\tau^-\,\overline{\nu})$ is predicted smaller than the SM
prediction while ${\cal B}(B_s\to\mu^+\mu^-)$ is larger than the SM
prediction. Precise measurements of ${\cal
  B}(B^-\to\tau^-\overline{\nu})$ at $B$ factories and ${\cal
  B}(B_s\to\mu^+\mu^-)$ at TEVATRON and LHC may play important roles
to search for signature of the MGM in addition to SUSY direct 
search at LHC.

The nature of the next-lightest SUSY particle is important in the
gauge-mediated models from viewpoints of collider physics and
cosmology. In Fig.~\ref{mstau-chi0} we show the lightest stau and
lightest neutralino masses as functions of $M_3$ for $N=1$ and $2$.
The stau mass is lighter than or quite degenerate with the neutralino
mass in the regions favored from ${\cal B}(\overline{B}\to X_s
\gamma)$ and $a_\mu$. This will have a important implication to SUSY
particles searches and also the studies at LHC and LC.

The parameter region $M_M\sim \Lambda$ is well-motivated from the
cosmological gravitino problem. When the $S$ field is directly coupled
with the dynamical SUSY breaking sector and $M_M\sim 100$~TeV,
gravitino may be lighter than $\sim 10$~eV. Such an ultra-light gravitino
is known harmless in cosmology \cite{Pagels:1981ke}.

\section*{Acknowledgments}

The works of J.H. is supported by the Grant-in-Aid for Science
Research, Ministry of Education, Science and Culture, Japan
(No.~19034001 and No.~18034002). Also that of Y.S. is supported by the
BK21 program of Ministry of Education, the KRF Grant
KRF-2005-210-C000006 funded by the Korean Government and the Grant
No.~R02-2005-000-10303-0 from the Basic Research Program of the Korea
Science and Engineering Foundation.

\setcounter{footnote}{0}


\begin{thebibliography}{99}

\bibitem{Dine:1981za}
   M.~Dine, W.~Fischler and M.~Srednicki,
  Nucl.\ Phys.\  B {\bf 189} (1981) 575;
%
  S.~Dimopoulos and S.~Raby,
  Nucl.\ Phys.\  B {\bf 192} (1981) 353;
%
   M.~Dine and W.~Fischler,
  Phys.\ Lett.\  B {\bf 110} (1982) 227;
%
  C.~R.~Nappi and B.~A.~Ovrut,
  Phys.\ Lett.\  B {\bf 113} (1982) 175;
%
  L.~Alvarez-Gaume, M.~Claudson and M.~B.~Wise,
  Nucl.\ Phys.\  B {\bf 207} (1982) 96;
%
  S.~Dimopoulos and S.~Raby,
  Nucl.\ Phys.\  B {\bf 219} (1983) 479.

\bibitem{Dine:1993yw}
  M.~Dine and A.~E.~Nelson,
  Phys.\ Rev.\  D {\bf 48} (1993) 1277
  [arXiv:hep-ph/9303230].

\bibitem{Dine:1994vc}
  M.~Dine, A.~E.~Nelson and Y.~Shirman,
  Phys.\ Rev.\  D {\bf 51}, 1362 (1995)
  [arXiv:hep-ph/9408384].

\bibitem{Dine:1995ag}
  M.~Dine, A.~E.~Nelson, Y.~Nir and Y.~Shirman,
  Phys.\ Rev.\  D {\bf 53}, 2658 (1996)
  [arXiv:hep-ph/9507378].

\bibitem{Dine:1996xk}
  M.~Dine, Y.~Nir and Y.~Shirman,
  Phys.\ Rev.\  D {\bf 55} (1997) 1501
  [arXiv:hep-ph/9607397].

\bibitem{Rattazzi:1996fb}
  R.~Rattazzi and U.~Sarid,
  Nucl.\ Phys.\  B {\bf 501} (1997) 297 
  [arXiv:hep-ph/9612464].

\bibitem{Gabrielli:1997jp}
  E.~Gabrielli and U.~Sarid,
  Phys.\ Rev.\ Lett.\  {\bf 79} (1997) 4752 
  [arXiv:hep-ph/9707546].

\bibitem{Gabrielli:1998sw}
  E.~Gabrielli and U.~Sarid,
  Phys.\ Rev.\  D {\bf 58} (1998) 115003 
  [arXiv:hep-ph/9803451].

\bibitem{Misiak:2006zs}
  M.~Misiak {\it et al.},
  Phys.\ Rev.\ Lett.\  {\bf 98}  (2007) 022002
  [arXiv:hep-ph/0609232].


\bibitem{Becher:2006pu}
  T.~Becher and M.~Neubert,
  Phys.\ Rev.\ Lett.\  {\bf 98}  (2007) 022003
  [arXiv:hep-ph/0610067].

\bibitem{Hagiwara:2006jt}
  K.~Hagiwara, A.~D.~Martin, D.~Nomura and T.~Teubner,
  Phys.\ Lett.\  B {\bf 649}  (2007) 173
  [arXiv:hep-ph/0611102].

\bibitem{Miller:2007kk}
  J.~P.~Miller, E.~de Rafael and B.~L.~Roberts,
  Rept.\ Prog.\ Phys.\  {\bf 70}  (2007) 795
  [arXiv:hep-ph/0703049].



\bibitem{Martin:1996zb}
  S.~P.~Martin,
  Phys.\ Rev.\  D {\bf 55} (1997) 3177
  [arXiv:hep-ph/9608224].

\bibitem{Bennett:2006fi}
  G.~W.~Bennett {\it et al.}  [Muon G-2 Collaboration],
  Phys.\ Rev.\  D {\bf 73}  (2006) 072003
  [arXiv:hep-ex/0602035].


\bibitem{Barberio:2007cr}
  E.~Barberio {\it et al.}  [Heavy Flavor Averaging Group (HFAG)
                  Collaboration],
  arXiv:0704.3575 [hep-ex].


\bibitem{Hall:1993gn}
  L.~J.~Hall, R.~Rattazzi and U.~Sarid,
  Phys.\ Rev.\  D {\bf 50}  (1994) 7048
  [arXiv:hep-ph/9306309].


\bibitem{Aubert:2004it}
  B.~Aubert {\it et al.}  [BABAR Collaboration],
  Phys.\ Rev.\ Lett.\  {\bf 93} (2004) 081802
  [arXiv:hep-ex/0404006].

\bibitem{Iwasaki:2005sy}
  M.~Iwasaki {\it et al.}  [Belle Collaboration],
  Phys.\ Rev.\  D {\bf 72} (2005) 092005
  [arXiv:hep-ex/0503044].


\bibitem{Huber:2005ig}
  T.~Huber, E.~Lunghi, M.~Misiak and D.~Wyler,
  Nucl.\ Phys.\  B {\bf 740} (2006) 105
  [arXiv:hep-ph/0512066].


\bibitem{Gambino:2004mv}
  P.~Gambino, U.~Haisch and M.~Misiak,
  Phys.\ Rev.\ Lett.\  {\bf 94} (2005) 061803
  [arXiv:hep-ph/0410155].



\bibitem{Scuri:2007py}
  F.~Scuri, f.~t.~CDF and D.~Collaborations,
  arXiv:0705.3004 [hep-ex].

\bibitem{Blanke:2006ig}
  M.~Blanke, A.~J.~Buras, D.~Guadagnoli and C.~Tarantino,
  JHEP {\bf 0610}  (2006) 003
  [arXiv:hep-ph/0604057].

\bibitem{Bobeth:2001sq}
  C.~Bobeth, T.~Ewerth, F.~Kruger and J.~Urban,
  Phys.\ Rev.\  D {\bf 64} (2001) 074014
  [arXiv:hep-ph/0104284].

\bibitem{Babu:1999hn}
  K.~S.~Babu and C.~F.~Kolda,
  Phys.\ Rev.\ Lett.\  {\bf 84}  (2000) 228
  [arXiv:hep-ph/9909476].



\bibitem{Isidori:2001fv}
  G.~Isidori and A.~Retico,
  JHEP {\bf 0111}  (2001) 001
  [arXiv:hep-ph/0110121].


\bibitem{Buras:2002wq}
  A.~J.~Buras, P.~H.~Chankowski, J.~Rosiek and L.~Slawianowska,
  Phys.\ Lett.\  B {\bf 546}  (2002) 96
  [arXiv:hep-ph/0207241].



\bibitem{Ikado:2006un}
  K.~Ikado {\it et al.},
  Phys.\ Rev.\ Lett.\  {\bf 97} (2006) 251802
  [arXiv:hep-ex/0604018].

\bibitem{Yao:2006px}
  W.~M.~Yao {\it et al.}  [Particle Data Group],
  J.\ Phys.\ G {\bf 33} (2006) 1.

\bibitem{Gray:2005ad}
  A.~Gray {\it et al.}  [HPQCD Collaboration],
  Phys.\ Rev.\ Lett.\  {\bf 95} (2005) 212001
  [arXiv:hep-lat/0507015].

\bibitem{Akeroyd:2003zr}
  A.~G.~Akeroyd and S.~Recksiegel,
  J.\ Phys.\ G {\bf 29} (2003) 2311
  [arXiv:hep-ph/0306037].

\bibitem{Hahn:2005cu}
  T.~Hahn, W.~Hollik, S.~Heinemeyer and G.~Weiglein,
{\it In the Proceedings of 2005 International Linear Collider Workshop (LCWS 2005), Stanford, California, 18-22 Mar 2005, pp 0106}
  [arXiv:hep-ph/0507009].

\bibitem{Pagels:1981ke}
  H.~Pagels and J.~R.~Primack,
  Phys.\ Rev.\ Lett.\  {\bf 48} (1982) 223;
  M.~Viel, J.~Lesgourgues, M.~G.~Haehnelt, S.~Matarrese and A.~Riotto,
  Phys.\ Rev.\  D {\bf 71} (2005) 063534
  [arXiv:astro-ph/0501562].

\end{thebibliography}
\end{document}